# Machine Learning in Detecting user's suspicious behaviour through Facebook wall

Aimilia Panagiotou[1], Bogdan Ghita[2], Stavros Shiaeles[2], Keltoum Bendiab[2]

[1] Open University of Cyprus, Faculty of Pure & Applied Sciences, 2220, Nicosia, Cyprus

[2] University of Plymouth, Centre for Security, Communications and Network Research, PL48AA, Plymouth, UK
sshiaeles@ieee.org

**Abstract.** Facebook represents the current de-facto choice for social media, changing the nature of social relationships. The increasing amount of personal information that runs through this platform publicly exposes user behaviour and social trends, allowing aggregation of data through conventional intelligence collection techniques such as OSINT (Open Source Intelligence). In this paper, we propose a new method to detect and diagnose variations in overall Facebook user psychology through Open Source Intelligence (OSINT) and machine learning techniques. We are aggregating the spectrum of user sentiments and views by using N-Games charts, which exhibit noticeable variations over time, validated through long term collection. We postulate that the proposed approach can be used by security organisations to understand and evaluate the user psychology, then use the information to predict insider threats or prevent insider attacks.

# I. INTRODUCTION

Our preferred methods of communication over the last years seem to have witnessed the highest speed of change in the whole human history [1]. In a short period of time, social media has become an integral part of our social life [1], [2]. Facebook, WhatsApp, Tencent QQ, Instagram, Twitter, Google, and LinkedIn are all examples of the rapid transfer of people interpersonal and social interactions from physical, human contact towards vast digital social communities. This transfer is happening on an unprecedented scale - Facebook alone currently has over 1 billion users and is expected to reach 1.69 billion registered accounts by 2020 [3], making it the most used social network worldwide. In light of the large amount of information provided by users about their personal life, such social networks become an increasingly significant public space that is used by organisations for data gathering, from academic research to business intelligence [1], [3], [4]. In this context, psychologists state that Facebook contains valuable indicators of mental health [1], and indeed the social media profiles of participants in several psychological tests seem to show some indicative content [5] and analysis of available data on Facebook can be used to produce user patterns, help to profile users and anticipate their future behaviour [6].

Facebook provides easy access to information provided by friends and partners, including changes to their profile [6], additions of new contacts, called friends, and messages posted on their virtual wall. This makes user information public within one's circle of friends, allowing their list of friends to be publicly harvested or exploited [6], providing a relevant platform for capturing behavioural attributes relevant to an individual's thinking, mood, communication, activities, socialisation and psychological changes [7], [8]. The emotion and language used in the messages posted on the user virtual wall may indicate feelings of loneliness, self-hatred, depression, and anxiety that, in turn, characterise mental illness [8], [9]. Inferring such relationships could also be relevant for national security organizations when trying to pre-empt malicious activities such as hiring hitmen, grooming the targets of paedophiles, or stealing identities [1].

Therefore, collecting and mining information of millions of users who are digitally expressing their feelings is of substantive value within many areas [1]. Generally, collection and manipulation of such information is done by Open Source INTelligence (OSINT) techniques [10]. Such techniques aim to discover useful "intelligence that is produced from publicly available information collected, exploited, and disseminated to an appropriate audience for the purpose of addressing a specific intelligence requirement", according to [10]. It differs from traditional intelligence techniques since it collects data from public sources such as social media, blogs and web communities [4]. OSINT can access to important personal information or links through social networks without any limitations [2]. To our best knowledge, OSINT and Facebook have not been used in any research to show the variations in user psychology over time. Thus, in this paper, we aim to contribute in this relevant field, by introducing a new approach that relies on machine learning and OSINT to detect alternations in the psychological profile of a Facebook user. The main contribution of this paper is the proposal of a novel machine learning method that allows information filtering, followed by the psychological user profile features mined in N-games layouts to feed the location finder framework proposed in [4]. In addition, the paper introduces a practical implementation for collection and processing of information from Facebook profiles, which may be stored in a database for further analysis. This research could be developed with more specific goals to serve other organizations; for example, it can be used by security organizations to detect and prevent malicious activities. It can also be used by public health to identify at-risk individuals or detect depression, and frame directions on guiding valuable interventions.

The paper is organised as follows: Section 2 describes the prior work discussing social media and its impact on individual feelings. In Section 3, we present the methodology of the proposed method using machine learning and OSINT to construct the information profile of a user. Section 4 presents experimental results and Section 5 provides concluding remarks and future work.

## II. TELATED WORK

Over the recent years, there has been growing interest in using social media as a tool for analysing emotions as expressed by people as part of their online interaction. In the context of Twitter, [12] highlighted that people publicly post information about their feelings and even their treatment on social media. Work in [13] studied linguistic and emotional correlation for postnatal changes of new mothers and built a statistical model to predict extreme postnatal behavioural changes using prenatal observations. This work highlights the potential of Twitter as a source of signals about likelihood of current or future episodes of mental illness. In a similar context, [9] demonstrated the potential of using Twitter as a tool for measuring and predicting major depression in individuals. The work proposed a variety of social media measures such as language, emotion, style, and user engagement to characterise depressive behaviour; supervised learning was used to construct classifier trained to predict depression, yielding 70% classification accuracy.

The authors of [14] focused on examining the relationship between social network, loneliness, depression, anxiety and quality of life in community-dwelling older people living in Dublin. In the study, thousands of community-dwelling people aged 65 and over were interviewed using the GMS-AGECAT diagnostic approach [15] and information from social networks was processed using the Practitioner Assessment of Network Type (PANT) schedule developed by Wenger [16]. The study found that loneliness and social networks independently affect the mood and wellbeing of the elderly, concealing a significant rate of underlying depression.

With the increasing societal importance of online social networks, their influence on the human collective mood state has become a matter of considerable interest and several academic research studies were undertaken. A long-term experiment was conducted over two years as part of [17] to study the influence of social networks on people's feelings. Several tests were performed on Facebook users, which consisted of manipulation of information posted on 689,000 user home pages, including news feeds, the flow of comments, videos, pictures and web links posted by other people in their social network. The study found that the mix of information could influence the overall mood of the user both positively or negatively through a process coined by the authors as "emotional contagion". Similarly [18] included a series experiments performed on social contagion, using several statistical approaches to characterise interpersonal influence with respect to a range of psychological and health-related characteristics such as obesity, smoking, cooperation, and happiness. Through the analysis of social network data, authors have suggested that the above characteristics do traverse and influence social networks; the study found evidence regarding social contagion in longitudinally followed networks. Work in [19], focused on studying the general happiness of Twitter users by recording and measuring their individual tweets. 129 million tweets were collected over a 6 month period, then a mathematical model was used to measure emotional variation of each user in time and how it is spread across links. The study found that general happiness, or subjective well-being (SWB), of these users is assortative across the Twitter social network. The study results imply that online social networks replicate social mechanisms that cause assortative mixing in a real social network. Thus, their ability to connect users with similar levels of SWB is an important factor in how positive and negative emotions are preserved and spread through human society.

Based on outlined research in this section, many experimental studies analysed the emotional variations of people over time and the influence of online social networks on these variations. In these studies, researchers have developed different approaches that can be used to measure user's emotions variation through social networks. All the performed experiments demonstrated the potential of using social media as a tool to study the mental status of users, but the majority of them were focused on the Twitter social network. In addition, most of them were experimentally conducted by using statistical approaches and in some cases with the physical presence of people under investigation. None of these approaches combines the simultaneous usage of a crawler for collecting information with a machine learning approach for filtering it and storing it in a database. Moreover, OSINT and Facebook were not applied by prior research to investigate the variations in user psychology. In this paper we propose a new approach that combines the above elements aiming to converge faster to more accurate results when compared to the studied approaches. The validation results indicate the proposed approach is effective and accurate.

## III. APPROACH OVERVIEW

As illustrated in Fig. 1, the first step consists in gathering relevant data from the Facebook social network and placing it in a database for further analysis. In this step, a web crawler has been developed and is used to collect the wall information from our Facebook friends. All these information were stored in a database for father analysis. The classification process is undertaken using the N-Gram technique. N-Grams are used for natural language processing to develop not just unigram models but also bigram and trigram models that will for developing features for our Learning Machine system. The Learning Machine system is where the data are further filter and the psychological profile of user is produced.

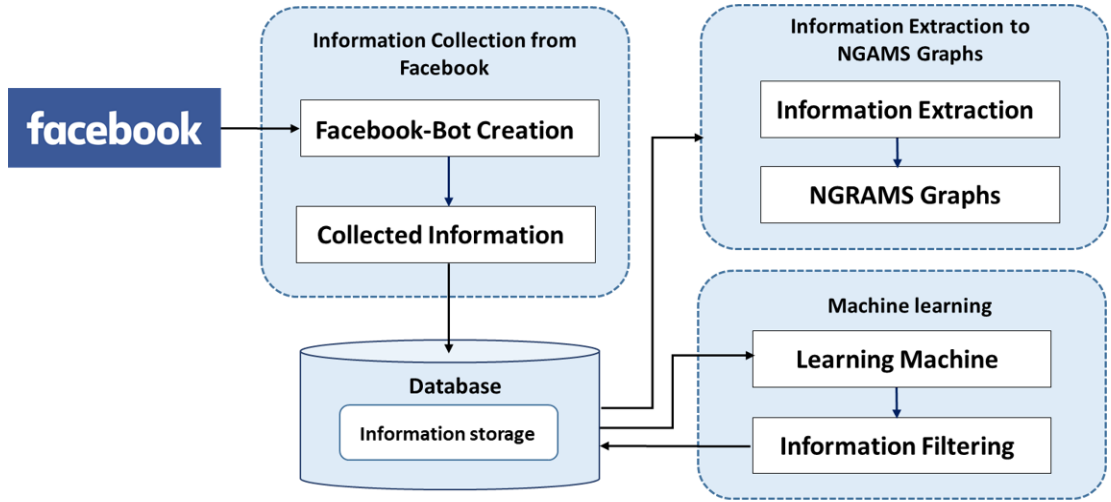

**Fig. 1.** Overview of the proposed method.

### 1) Information collection from Facebook

On Facebook, people frequently express and share their negative and positive emotions [20], which are later seen by their friends via a "News Feed" page [21], which dynamically updates the activities of the Facebook friends of a user [21]. Authors in [20] confirmed that people use News Feed every day to stay in touch with friends and family, and to stay informed about the world around them. It is worth nothing that, in this study, information was initially collected from the content shared by friends of the research team, then it was expanded to friends of friends, with their consent. The data was gathered via a web crawler and stored in a database (see Fig. 1). In the initial stage of the data gathering process, a Facebook Bot [22] was created for automatic collection and storage of each user available information including posts (the content and the associated timestamp), personal details (name, Facebook URL, gender) and photos. The Facebook crawler uses Selenium [23] to automatically connect to Facebook and propagate to the news feed to scarp the HTML at that URL to gather, cache and extract the aforementioned information.

### 2) Information extraction and classification

The information collected from profiles of friends through Facebook Wall are sent to the machine learning components for pruning the posts from irrelevant content such as, URL links, Facebook user names and articles ("a", "an", "the"). Then, stored the result in the Database (see **Fig. 1**). The results stored in the database are used in the process of the N-grams graphs construction, where the information related to user psychology are mined using N-Grams layouts. The Python class NGrams [24] is used for the N-Grams graphs construction.

In this paper, we use English langue, however, our approach can be updated to other languages. We assume that an emoticon within a post represents an emotion for the whole post and all the words of this post are related to this emotion. Thus, we split the psychological profiles in four classes: "Happy", "Sad", "Love" and "Disappointment".

- **Happy**: this class present word that design filing of happiness such as "happy", ":-)", ":)", "=)", ":D" etc.
- **Sad**: this class represent words that present emotion of sadness such as "sad", the symbols'☹' ":-(", ":(", "=(", ";(" etc.
- **Love**: this class consists of words that present feeling of love such as "love", '<', '3', "<3", etc.
- **Disappointment**: this class consists of words that present feeling of frustration such as "disappointed", "anger", etc.

## IV. EXPERIMENTAL RESULTS ABD ANALYSIS

### 1) Experimental setup

In this section, we present the analysis results of the implemented prototype, based on the methodology from section 3. The experiments were performed on a virtual machine with an i5 CPU and 8GB of RAM, using a Python-based software implementation. For the creation of Facebook_Bot as well as the FaceWallGrap application, Python programming language was used due its extensibility and wide library support. We have tested our approach on a set of real Facebook posts (see Fig. 2.) in time period from 2010 until 2016.

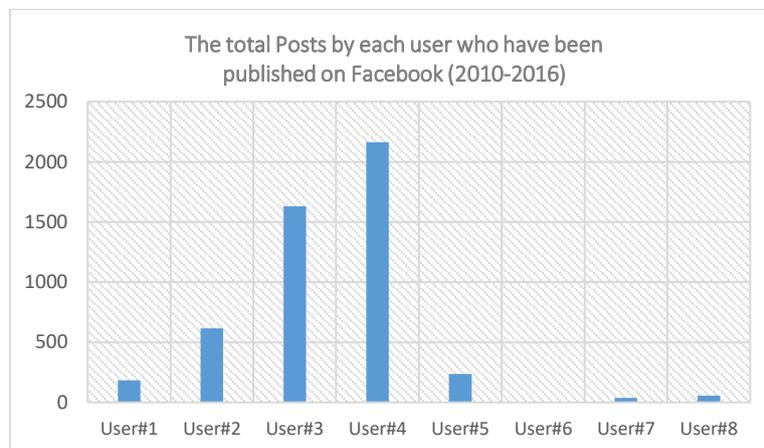

**Fig. 2.** The total Posts by each user who have been published on Facebook over the period (2010-2016)

### 2) Results analysis

The following graphs summarise the results of our Facebook experiment using our Facebook account wall. As illustrated in Fig.3 (a), The "Happy n-gram" shows the feeling of joy in time periods from a total result for all users. The words used to create this n-gram are words and symbols that belongs to the class "Happy". It seems that in this graph statistically in the past 2-3 years users have ups and downs in the subject of joy while in the past years they were stagnant. The "Sad n-gram" graph (Fig.3 (b),) shows the feeling of regret in time periods from a total result for a user. The words used to create this n-gram are words and symbols that belongs to the class "Sad". Moreover, it seems that in this graph statistically the last two years users have very few and specific moments that are sad. The Love n-gram graph (Fig.3 (c)) shows the feeling of love in time periods from a total result for a user. The words used to create this n-gram are words, numbers and symbols that belong to the class "Love". Additionally, it seems that in this graph statistically the 2-year years users have increased moments who are in love or love their round. Whereas, the graph "Disappointed n-gram" (Fig.3 (c)) shows the feeling of frustration over time by a total result for all users. The words used to create this n-gram are from the class "Disappointed" that reveals disappointment. In addition, it seems that in this graph statistically the last 3 years users have increased moments that are disappointed. The graph "Disappointed n-gram" (Fig.3 (d)) shows the feeling of frustration over time by a total result for all users. The words used to create this n-gram are from the class "Disappointed" that reveals disappointment. In addition, it seems that in this graph statistically the last 3 years users have increased moments that are disappointed.

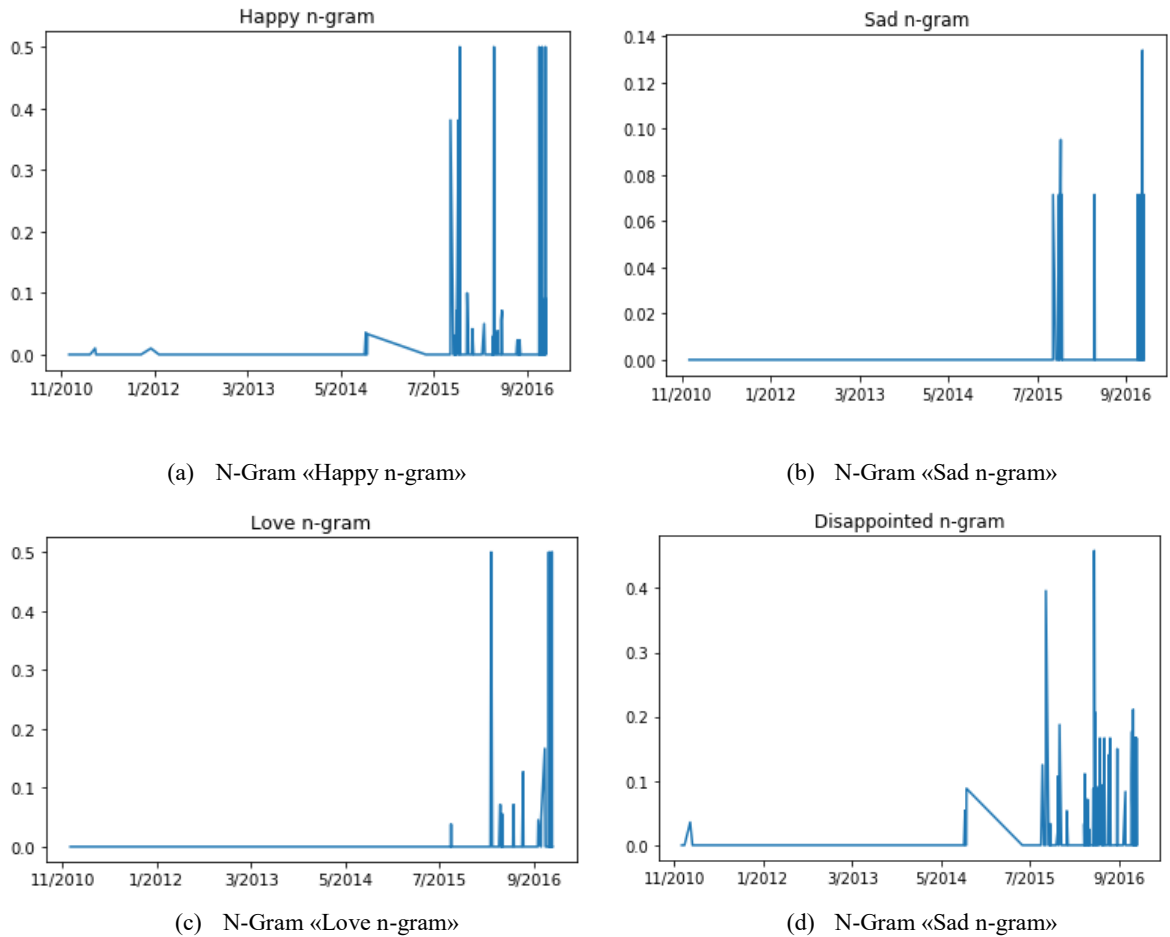

(a) N-Gram «Happy n-gram»

(b) N-Gram «Sad n-gram»

(c) N-Gram «Love n-gram»

(d) N-Gram «Sad n-gram»

**Fig. 3.** n-gram graphs for the Facebook accounts wall

## V. CONCLUSION

The purpose of this paper was to extend the system proposed in [4] by adding a suspicious user classification component based on psychological profiling. It can be concluded that social networks such as Facebook may be used, through n-gram extraction, for user profiling and it is also noteworthy that this procedure is complete transparent to the user. It can be also concluded that certain environmental and social context conditions are important factors and may affect emotions. Future extension of this work would be the integration with the proposed model in [4], the addition of more psychological profiles, retrieving information from other social networks such as Twitter, Instagram and finally process pictures uploaded from users and classify their emotion based on face detection. This combined framework could be proven beneficial for law enforcement and be able to prevent unwanted situations.